# Assessing the regulatory framework of financial institutions in Canada in the context of international climate risk management practices and Canadian net zero emission targets

*Víctor Cardenas\**

## Abstract

The document addresses the essential role of financial regulatory frameworks in mitigating climate-related risks within the financial sector. The assessment evaluates Canada's efforts to establish a regulatory framework for financial climate risk disclosure, compares it to international standards, and identifies areas for improvement. The regulatory framework, fiscal impact, and incentives to relinquish investments in fossil fuel projects and transition to renewable energies swiftly and seamlessly are the primary challenges that Canada faces in achieving energy net zero targets.

\*   Institute for Resources, Environment and Sustainability (IRES), UBC

## TABLE OF CONTENTS

## 1. Executive Summary

Climate change has the potential to threaten financial institutions in each country as well as global financial stability. Supervisory agencies (FSB, 2022) have an enormous responsibility ensure that climate change risks do not undermine the stability of their respective financial systems.
These financial climate risks, commonly referred to as "climate-related risks," can be broadly categorized as physical and transition risks (TFCD, 2021). Physical risks pertain to financial vulnerabilities arising from the increasing severity and frequency of climate-related extreme events (referred to as acute physical risks), the gradual and long-term shifts in climate patterns (known as chronic physical risks), and the indirect consequences of climate change, including impacts on public health, such as morbidity and mortality effects On the other hand, "transition risks" are financial risks linked to the process of adapting to a low-greenhouse gas (GHG) economy. These risks can arise from existing or future government policies, regulations, and legislation aimed at curbing GHG emissions. They can also be influenced by technological advancements, shifts in market dynamics, and changes in customer sentiment towards a low-GHG economy.

Accordingly, the financial supervisors must take a future-directed approach and consider these risks as real threats to the financial systems they regulate (OECD, 2015). However, the level of readiness of the regulatory frameworks to regulate these risks remain a challenge.

The Intergovernmental Panel on Climate Change (IPCC) found that each of the last four decades has been successively warmer than any decade that preceded it since 1850. Global surface temperature in the first two decades of the 21st century (2001-2020) was 0.99º Celsius (C) higher than 1850-1900.

The Assessment Report VI (Calvin et al., 2023) published by the IPCC provides overwhelming evidence that human activity is rapidly affecting global climate and contributing to the warming of the planet. This upward trend in temperature, along with other shifts in precipitation and extremes will affect economic activity and the financial sector more generally.

According to Climate Initiative Policy (Buchner et al., 2023), global warming could cause losses in the business as usual (BAU) scenario between 2025-2100 to the tune of USD 2,328 trillion. Some studies[1] indicate that an increase in warming above 3ºC could lead to global losses of 18% of GDP by 2050 and 20% by 2100.

However, there are currently no accurate estimates of the size of losses and future losses for individual firms, and only estimates by sector are available. It is precisely this gap in the data that must be addressed, with an analysis of the concerns for financial markets players, regulators, and policy makers.

In this context, regulatory frameworks have emerged as the first formal and articulated response to address the challenge of climate change for economies and financial systems (FSB & NGFS, 2022). The primary goal of these frameworks is to ensure that the regulator and the markets have relevant information from companies on their emissions, the degree of exposure to climate risk, and their internal policies for managing climate risk.

This report contextualizes the Canadian financial authority effort, in order to add to the discussions of regulatory frameworks for climate risk related disclosures and considers similar efforts by its peers in the G7 group of economies. Our assessment concludes that Canada is developing a regulatory framework (OSFI, 2023a) which is consistent with its peers.

---

[1] Swiss Re Institute. 2021. The economics of climate change: no action not an option. Available at: (link does not work)
https://www.swissre.com/dam/jcr:e73ee7c3-7f83-4c17-a2b8- 8ef23a8d3312/swiss-re-institute-expertise-publication-economics-of-climate-change.pdf

However, several areas of opportunity have been identified. For example, the Standardized Climate Scenario Exercise (SCSE), which is the current approach of choice, focuses more on transition risk than physical risk. It also seems to highlight the role of the banking sector while paying less attention to the insurance sector. It is not clear if the purpose of this limited scope is to make the delivery of a feasible regulatory framework in the coming years. If so, it would be helpful to know if a long-term strategy to fully cover both risks and both sectors is envisioned. In the future, SCSE will shape the regulation; therefore, it is crucial not to skimp on any details.

We concur that the current approach shown in the SCSE is achievable in the short term. However, in the medium and long term, physical risk should be part of the assessment with the same level of detail. A survey of twenty-six Network for Greening the Financial System (NGFS) (NGFS, 2023) members this year found broad consensus on the urgency of considering compound risks when analyzing climate change impacts, and part of that compounding risk is the physical risk.

The relevance of the financial sector to the economy is crucial, particularly in the context of net zero emissions targets for 2050. The challenge facing the financial regulator is unprecedented, demanding an extraordinary amount of technical effort. In this context, the financial regulator could strengthen its team of experts, particularly those with a focus on studies of climate change and the financial sector.

The aim of this report is to provide an assessment of current trends among G7 regulatory authorities in regulation of climate-related risk. Considering regulation on climate related-risk disclosure developed by its peers and international standards issued in that topic, this report assesses the current regulatory proposal by Canadian financial authorities and serves to add to the ongoing dialogue among market players[2], academia (Galvez Rosa et al., 2022) and authorities (CESD, 2023).

---

[2] https://www.pwc.com/ca/en/today-s-issues/environmental-social-and-governance/net-zero/tcfd-reporting-requirements.html

## 2. Canada's fossil fuel financing

According to the Canadian Association of Petroleum Produers (CAPP) (2023), oil and natural gas production in Canada ranks fourth and fifth largest globally, respectively.

Canada, with a rich 160-year history in the oil sector, has established itself as a frontrunner in developing cutting-edge technologies for extracting petroleum and refining its byproducts. Canada's oilfields are characterized by the presence of hydrocarbons that include a diverse range of product types, ranging from low-density gaseous molecules such as natural gas, to high-density sticky bitumen from oil sands.

Crude oil sands account for approximately 58% of Canada's total oil production. The oil sands formation produces an unusually viscous petroleum known as bitumen. Conventional production refers to approximately 38% of Canada's crude and nearly all its natural gas. Oil and gas were historically extracted from vertical wells; however, hydraulic fracturing and horizontal wells are currently the predominant methods. Approximately 4% of Canada's oil production is derived from four offshore developments.

From a financial standpoint, the oil industry is distinguished by its high capital intensity and lengthy maturity periods, which are necessary to realize benefits. In general, all previous studies and preparations for the exploitation of a field, as well as, in particular, to be linked to specific tax frameworks designed for such sector, as well as, once the proven and potential reserves are confirmed, the investment in infrastructure necessary for its exploitation. Thus, it is an investment that necessitates including such expenses to substantiate its profitability. It is a common practice for certain valuation methodologies to advise against making such investments, even if there are proven reserves, if the market prices do not justify that investment.

The oil and gas annual revenue for 2024 is currently estimated to be $188 billion, as oil prices have experienced a significant increase since the beginning of the year. The economic implications of Canada's upstream oil and gas sector are substantial. Over 3% of Canada's GDP was accounted for by the sector in 2023. In Canada, the Oil and Gas Extraction sub-industry is the greatest industrial producer of products. It is 20% larger than the engineering and other construction activities sub-industry, which is the second largest, and 30% larger than the residential building construction industry (Statistic Canada). Finally, the industry paid a record $33 billion in oil and gas royalties to provincial governments in 2022. In 2023 and 2024, over $20 billion is expected in each year. This industry and derivative products are also a critical component of Canada's exports, making up about 20% of Canada's balance of trade.

Since its inception, the oil and gas industry has grown in tandem with the financial sector, which provides the capital required to finance the exploitation of deposits, as a result of the capital-intensive character of the industry. Consequently, the financial industry is an essential component of the oil and gas industry supply chain; without it, the oil industry would not be able to sustain its development and dynamism.

According to Hudson and Bowness (2021), the financial sector is deeply intertwined with the fossil fuel sector in Canada. Major Canadian financial firms have significant investments in the extraction and transportation of oil and gas, with a substantial portion of their portfolios dedicated to energy, particularly oil and gas extraction and transportation. This indicates a strong interest in the future profitability of fossil fuel firms. Additionally, the financial industry plays a vital role in providing capital for fossil fuel projects, as these projects are capital-intensive and rely on debt or equity finance for development. Furthermore, financial institutions also provide insurance for extraction and transportation activities in the fossil fuel sector.

The Canadian financial sector features five major banks, The Royal Bank of Canada (RBC), Canadian Imperial Bank of Commerce (CIBC), Scotiabank, Toronto Dominion (TD) and the Bank of Montreal (BMO). These banks provided CAD $ 71 billion between 2015 to 2017, it is hard to estimate the total financing to oil and gas industry, because there is not full disclosure of this data, despite rough estimations by research funded by activist organizations.

From different perspective on the same topic, the Bank of Canada and the Office of the Superintendent of Financial Institutions (OSFI) developed a study assessing total financial system climate-relevant exposure, this includes direct and indirect interconnects with the whole supply chain of oil, gas and derivates, based on data from 2022, and assessing all regulated financial entities, including deposit-taking institutions (e.g. banks), insurance, pensions and investments funds, in total the financial system assessed is about CAD$ 14.4 trillion, in the following financial assets: loans and private debt, bonds, public equity and private equity (for private funds only).

The outcome of the study shows banks have climate-relevant exposures; deposit-taking institutions have under 4% exposure to climate-relevant assets (CAD$ 7.9 trillion total assets under management, AUM), while life insurance companies have about 19% (CAD$0.9 trillion total AUM), pensions funds 15% (CAD$ 2.7 trillion of total AUM) and investments funds 11% (CAD$ 2.9 trillion of total AUM). In total, CAD$ 1.2 trillion directly or indirectly finances the oil and gas industry. This implies that around 8.4% of the Canadian financial system is exposed to this industry.

Despite its commitment to a net zero-emissions economy by 2050, the government is confronted with a significant challenge in its role as the primary champion in addressing climate change. Given the complex and profound relationship between the oil and gas industry, which not only contributes to tax revenues but also creates and distributes wealth in the economy, such a challenge requires a deep transformation of the economy. Consequently, a transformation of this nature calls for a change in the financial system. The challenge is twofold: how to achieve a net emissions target without disrupting with the economy's growth path.

One of the primary objectives of this transition to a net-zero emissions economy is to reorganize the regulatory framework for the financial sector to highlight the climate related risk as a central piece of the regulation as well as the financial risk currently regulated. The government agencies responsible for the study, design, and issuance of these standards have been engaged in this issue. The subsequent sections evaluates whether there are opportunities for improvement in this process concerning climate related risk.

Canada has established a target of decreasing emissions to 40%–45% below the levels recorded in 2005 by the year 2030. Additionally, Canada aims to achieve net-zero emissions by 2050, with the specific purpose of limiting global warming to 1.5°C, as stated by the International Institute for Sustainable Development (IISD). However, the 2030 target falls well short of the 60% reduction below 2005 levels that Canada is obligated to accomplish in order to meet its commitment (Holz et al., 2019). The International Energy Agency (IEA) has produced a comprehensive scenario called "Net Zero Emissions by 2050" on a worldwide scale (IEA, 2021). As to Greenpeace et al. (2022), this situation requires a yearly reduction of 3%–4% in the output of oil and fossil gas, without any new oil or fossil gas fields being created.

According to the International Institute for Sustainable Development (IISD), based on the principle of common but differentiated responsibility and with a focus on fairness, recent research has concluded that Canada needs to exceed its current targets and decrease oil and gas production by 74% by 2030 to meet its obligations as a high-emitting nation (Calverley & Anderson, 2022). However, according to the national emissions inventory, Canada's emissions have only fallen by 1.1% during the previous 15 years (prior to the pandemic). Additionally, the oil and gas industry remains the sector with the highest rate of growth in the Canadian economy (Al-Aini et al., 2022).

Amounts provided by the Canadian financial system, around 8% of AUM, are directly and indirectly related to the oil and gas industry. If the decreasing emissions targets requires a yearly reduction of 3%–4% in the output of oil and fossil gas, then the financial assets exposed should be reabsorbed smoothly and gradually by the rest of the system to another sector without compromising the integrity of the financial system or the economic growth path. A regulatory framework provides guidance precisely to avoid financial instability, slowdowns in the economy and specifically a drop in quality of life for populations.

## 3. Climate finance: overview on current public policy trends

### 3.1. Climate related risk regulation global framework

The concern of a real threat of climate change on the financial system was made public in a September 2015 speech by the Governor of the UK Central Bank "Breaking the tragedy of the horizon - climate change and financial stability"(Carney et al., 2015). The speech highlighted that climate risk is not only a scientific concept but actually comes with financial and business risks as well.

Tragedy is already here. The world's recent disasters, including unprecedented floods in Pakistan, forest fires in Canada and the US (the worst in decades), heatwaves in Europe, and hurricane Otis in Mexico (the strongest ever recorded in the Western Pacific) among many others, show that the negative effects of global warming effects are occurring earlier than scientific experts (Calvin et al., 2023) initially thought.

Climate change poses significant risks to the financial sector globally. For instance, studies (Nie et al., 2023) suggest that severe climate and environmental disasters can lead to an increase in non-performing loans (NPLs) within the banking sector. Other financial players like insurance companies, asset managers, and stock and debt issuers are also vulnerable to climate-related risks. For example, climate events can lead to physical damages, economic losses, and disruptions that impact financial institutions' solvency and profitability. As a result of climate change, default risk can be increased (Iqbal & Nosheen, 2023), which is the possibility of financial entities going bankrupt. This could jeopardize global financial stability (FSB, 2022).

Mark Carney's Bank of England Governor speech in 2015 prompted a change of perspective on the potential impact of climate change on the financial system, which had been long underestimated. Until that point, no relevant data had been disclosed, nor were there any metrics, methodologies, estimates of such impacts, or more importantly, quantifications and directions of such impacts. Conceptually, it was clear that climate change is a threat to humanity and economies will be affected. As a core part of those economies, financial systems will be impacted as well.

In 2015, the first response was through the Financial Stability Board[3] ("FSB"), an international organization composed by financial regulators from G20 countries and multilateral organisms, whose primary objective is to promote international financial stability by coordinating and monitoring the implementation of financial regulatory reforms and best practices. The FSB assumed the responsibility of refocusing financial regulation, and considering the potential hazard of climate change to the global financial system. By the end of 2015, the FSB promoted the creation of the Task Force on Climate-Related Financial Disclosures (TFCD)[4].

---

[3] (https://www.fsb.org/about/#mandate

TFCD (Carney, 2017) originally aimed to develop voluntary climate-related financial risk disclosures for financial institutions and non-financial corporations. Investors and other members of the public would benefit from these disclosures if companies adopted them. It assumed that financial market participants are capable of utilizing, processing, and interpreting this information to assess each financial institution's climate risk. The TFCD disclosure recommendations are structured around four thematic areas that represent core elements of how organizations operate: governance, strategy, risk management, and metrics and targets. The design and release of recommendations concerning how and what to disclose in terms of relevant information to estimate the degree of exposure to climate risk, as well as the number of adherents to such recommendations, has been ongoing.

In the period 2016 to 2021, many other initiatives emerged promoting the disclosure of information showing the degree of exposure to climate risk. These include the CDP questionnaire[5], SASB standard[6], GRI standard[7], Integrated Reporting Framework[8], and the CDSB framework[9]. In 2022, many of the initiatives mentioned, including the TFCD (where the disclosure methodology was born), adhered to the International Financial Reporting Standards (IFRS) foundation. The IFRS took over the monitoring of climate-related disclosures from the TFCD[10] and is slated to become the global champion for issuing guidelines in terms of disclosure of information on climate risk exposure.

While some of the other standards continue to exist on their own, the authorities representing the FSB have given that mandate to the IFRS Foundation, such that the standard to be followed is now given by two new disclosure guidelines: the IFRS S1 on sustainability issues and IFRS S2 (ISSB, 2023b) on climate related risk. The disclosure of information, however, is still only a first step towards estimating climate risk. Regulators and financial markets are still working on how to assimilate the information disclosed to authorities and the markets and then assess climate risk based on methodologies that are technically acceptable to the international community.

The Paris Agreement (UNFCCC, 2015) heralded important changes in the global financial sector. Since then developed economies have established initiatives aimed at identifying climate risk as a real threat to the global financial system. Within each government and multilateral institution, multidisciplinary teams have been strengthened, incorporating the environmental and climate approach from the financial perspective.

The initiatives for estimating climate risk in corporations and the financial systems are not limited to the government's efforts. Several companies and organizations have emerged that propose methodologies for the analysis of information through already publicly disclosed information, as well as private company information. However, the degree of consensus that exists on the type of information disclosed is not the same as on the methodologies for analyzing such information. These independent methodologies have their own climate risk definitions, which may or may not be compatible with the regulatory framework still in discussion.

---

[4] https://www.fsb-tcfd.org/
[5] https://www.cdp.net/en
[6] https://sasb.org/standards/
[7] https://www.globalreporting.org/standards/
[8] https://www.integratedreporting.org/resource/international-ir-framework/
[9] https://www.cdsb.net/what-we-do/reporting-frameworks
[10] https://www.ifrs.org/news-and-events/news/2023/07/foundation-welcomes-tcfd-responsibilities-from-2024/

Despite the progress, significant challenges still lie ahead. For example:

- There are still many doubts regarding the how and what to disclose particularly in specific niche industries or subsegments of businesses that are subject to the proposed regulation.

- Even though the information disclosed would be available in accordance with the proposed regulations, it is unclear what methodologies should be used by regulators, investors, clients, and suppliers to obtain comparable and relevant information.

- Climate scenarios for the financial sector require further study in order to be applied properly for valuation and analysis of companies. There is still a long way to go until the entire financial sector is aware of the importance of these types of scenarios.

- Financial entities' shareholders remain concerned about potential changes in the capital requirements of financial institutions based on climate risk assumed since the methodology for estimating climate risk is still under consideration.

- Risk management methodologies are at a breaking point. There is a current debate in academia on how to methodologically incorporate climate risk, as a financial risk, into the vast risk portfolio that a company manages.

- Systematic and orderly disclosure of non-financial information is an issue that many economies are still not addressing in their priority agenda.

| **Box 3: TFCD approach and recommendations** |
|---|
| The TFCD structured its recommendations around four thematic areas that represent core elements of how organizations operate: governance, strategy, risk management, and metrics and targets (Figure 1). The four recommendations are supported by recommended disclosures that build out the framework with information that will help investors and others understand how reporting organizations assess climate-related risks and opportunities. <br><br> Figure 1 <br>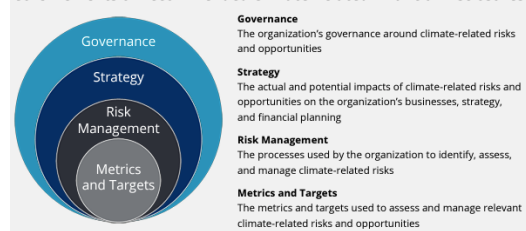<br> The TFCD framework takes a risks and opportunities approach. The breakdown is as follows: 1) transition risk, which includes policy and legal, technology, market, and reputation risks, and 2) physical risk, assessed in acute and chronic risk. Opportunities assess resource efficiency, energy sources, product/services, markets, and resilience. Both assessments should fall in cascade over the strategic planning and risk management of the entities assessed, and then analyze the financial impact, which should include the flow of revenues and expenditures and on the other side, the stocks shown on assets & liabilities and capital & financing. In between, the flows and stocks are reflected in the income statement, cash flow statements, and balance sheets. (see figure 2). |

Figure 2

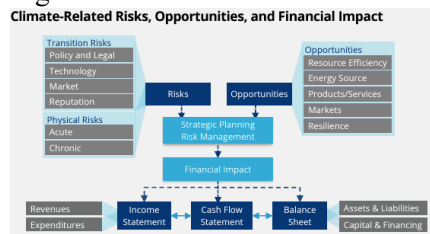

The main recommendations are in the four thematic disclosures in terms of governance, strategy, risk management, and metrics and targets, see figure 3.

Figure 3

## Recommendations and Supporting Recommended Disclosures

| Governance | Strategy | Risk Management | Metrics and Targets |
|---|---|---|---|
| Disclose the organization's governance around climate-related risks and opportunities. | Disclose the actual and potential impacts of climate-related risks and opportunities on the organization's businesses, strategy, and financial planning where such information is material. | Disclose how the organization identifies, assesses, and manages climate-related risks. | Disclose the metrics and targets used to assess and manage relevant climate-related risks and opportunities where such information is material. |
| **Recommended Disclosures** | **Recommended Disclosures** | **Recommended Disclosures** | **Recommended Disclosures** |
| a) Describe the board's oversight of climate-related risks and opportunities. | a) Describe the climate-related risks and opportunities the organization has identified over the short, medium, and long term. | a) Describe the organization's processes for identifying and assessing climate-related risks. | a) Disclose the metrics used by the organization to assess climate-related risks and opportunities in line with its strategy and risk management process. |
| b) Describe management's role in assessing and managing climate-related risks and opportunities. | b) Describe the impact of climate-related risks and opportunities on the organization's businesses, strategy, and financial planning. | b) Describe the organization's processes for managing climate-related risks. | b) Disclose Scope 1, Scope 2, and, if appropriate, Scope 3 greenhouse gas (GHG) emissions, and the related risks. |
| | c) Describe the resilience of the organization's strategy, taking into consideration different climate-related scenarios, including a 2°C or lower scenario. | c) Describe how processes for identifying, assessing, and managing climate-related risks are integrated into the organization's overall risk management. | c) Describe the targets used by the organization to manage climate-related risks and opportunities and performance against targets. |

*Source: Taken from Final Report: "Recommendations of the Task Force on Climate-related Financial Disclosure, TCFD, 2017*(TFCD, 2021).

*Current state of the TFCD recommendations*

Currently, the progress in the implementation of the TFCD recommendations is assessed in terms of voluntary disclosures made by companies on an individual basis. At the national level, many economies, mostly developed ones, are in the process of drafting regulatory frameworks for non-financial information disclosure consistent with TFCD recommendations.

The TFCD (2023) report (TFCD, 2023) included a review of 1,350 large, geographically diverse, and stock market listed companies reports over a three-year period (from 2020 to 2022). The main findings are summarized in the following table:

| Box 1: Main findings |
|---|
| • The percentage of companies disclosing TCFD-aligned information continues to grow, but more progress is needed. For fiscal year 2022 reporting, 58% of companies disclosed information in line with at least five of the 11 recommended disclosures — up from 18% in 2020; however, only 4% disclosed in line with all 11. |
| • The percentage of companies reporting on climate-related risks or opportunities, board oversight, and climate-related targets increased significantly — by 26, 25, and 24 percentage points, respectively — between fiscal years 2020 and 2022. |
| • Disclosure of climate-related financial information in financial filings is limited. On average for fiscal year 2022, information aligned with the 11 recommended disclosures was four times more likely to be disclosed in sustainability and annual reports than in financial filings. |
| • In 2022, the most often disclosed recommended disclosure — at 71% of the companies reviewed — was the metrics they use to assess their climate-related risks or opportunities. |
| • The least disclosed recommended disclosure for all three years reviewed was the resilience of companies' strategies under different climate-related scenarios, with only 11% disclosing this information in 2022. |
| Source: Taken from Task Force on Climate-related Financial Disclosures 2023 Status Report |

Considering that such disclosures have been made on a voluntary basis, the results are encouraging. However, there is a clear need for regulatory frameworks to enforce mandatory disclosure of non-financial information, given the need for prompt results consistent with the climate emergency.

On national efforts to regulate and estimate climate risk, the results are mixed. The outcomes of such efforts can be split in two categories:
(i) a new set of climate related risk disclosure regulatory frameworks proposed by country, and
(ii) design, definition, and development of hypothetical scenarios, explicitly prepared for fiscal and financial policy making. These scenarios provide a common and up-to-date reference point for understanding how climate change (physical risk) and climate policy and technology trends (transition risk) could evolve in different futures.

The G7 countries[11] are primarily involved in the development of climate related risk disclosure regulatory frameworks . Through their central banks and financial supervisors, these nations are developing regulatory drafts or regulations considering TFCD recommendations on non-financial information disclosure on a mandatory basis.

Central banks and financial supervisors are increasingly undertaking scenario analysis to identify, assess, and understand how to mitigate climate risks in the financial system. The Network of Greening the Financial System (NGFS) has been the main vehicle to develop this kind of analysis (NGFS, 2022). The membership of this global association grew from 8 central banks and supervisors committed in 2017 to 121 countries and 19 observers (multilateral organisms) by the end of 2022.

| Box 2: Comparison among NGFS, IPCC and IEA scenarios |
|---|

---

[11] Canada, France, Germany, Italy, Japan, the United Kingdom, and the United States, as well as the European Union.

> The Network for Greening the Financial System (NGFS) scenarios have been developed to provide a common starting point for analyzing climate risks to the economy and financial system. They have been created as a tool to shed light on potential future risks and to prepare the financial system for shocks that may arise. It is important to emphasize that the NGFS scenarios are not forecasts. Instead, they aim to explore the bookends of plausible futures (neither the most probable nor desirable) for financial risk assessment. The NGFS scenarios were developed by a consortium from the Potsdam Institute for Climate Impact Research (PIK), International Institute for Applied Systems Analysis (IIASA), University of Maryland (UMD), Climate Analytics (CA) and the National Institute of Economic and Social Research (NIESR).
>
> - The NGFS Phase II scenarios were assessed in AR6 of IPCC WG III and were integrated in its final report. They cover a small range of input and model assumptions but have, on average, higher sectoral and regional granularity than the rest of emission scenarios assessed by WG III.
> - The NGFS Phase II GCAM Current Policies scenario was selected as a Reference pathway by the WG III as a comparison to the Illustrative Mitigation Pathways.
> - Because they were developed for risk assessment purposes, the NGFS Scenarios do not always have equivalents in the IEA or IPCC realms; the latter focuses on exploring transition pathways. However, Net Zero 2050 scenarios are well aligned on several dimensions.
> -
>
> Source: Taken from NGFS Scenarios for central banks and supervisors September 2022.

*The future of TFCD recommendations*

According to the FSB, the duties assigned to the TCFD in regard to the development of the climate related framework has culminated in its mandate. At the same time, the International Sustainability Standards Board (ISSB) of its climate-related and general sustainability-related disclosure standards was released, assuming that mandate in its standard, which are strongly based in the TCFD recommendations.

According to the International Financial Reporting Standards (IFRS) Foundation, once they released ISSB Standards—IFRS S1(sustainability norms) and IFRS S2 (climate related norms)—the FSB has asked the IFRS Foundation to take over the monitoring of the progress on companies' climate-related disclosures from the TCFD.

The ISSB through the IFRS Foundation released two set of norms: 1) IFRS S1 provides a set of disclosure requirements designed to enable companies to communicate to investors about the sustainability-related risks and opportunities they face over the short, medium and long term, and 2) IFRS S2 sets out specific climate-related disclosures and is designed to be used with IFRS S1.

Regarding IFRS S2 (ISSB, 2023a) norms which are linked specifically with the TCFD recommendations, the aim of the norm is to "*require an entity to disclose information about its climate-related risks and opportunities that is useful to users of general-purpose financial reports in making decisions relating to providing resources to the entity.*"

In accordance with IFRS S2, entities are required to disclose information regarding the risks and opportunities related to climate change that could have a significant effect on their cash flows, their access to credit, or their cost of capital in the short, medium, or long term (known collectively as 'climate-related risks and opportunities' which may affect an entity's prospects in the short, medium or long run).

The IFRS S2 follows the same structure in terms of the TCFD recommendations, considering the climate risk breakdown in terms of physical risk and transition risk, as well as the four themes of governance, strategy, risk management, and metrics.

IFRS S2 standards were issued in 2023 and serve as a mandatory guide for all countries that adhere to the standards issued by the IFRS foundation. In other words, this is a first step for each country, through its supervisors or financial authorities, to apply this international standard. In turn, each authority will have to issue specific regulations based on the international standard for the application of the same. Therefore, countries that have not yet developed a standard in this direction still have a long way to go following IFRS S2.

### 3.2. Climate finance commitments

According to the United Nations Framework Convention on Climate Change[12] (UNFCCC), climate finance refers to "*local, national or transnational financing—drawn from public, private and alternative sources of financing—that seeks to support mitigation and adaptation actions that will address climate change.*"

In accordance with the principle of "common but differentiated responsibility and respective capabilities" set out in the Convention, developed country Parties are to provide financial resources to assist developing country Parties[13] in implementing the objectives of the UNFCCC.

| Box 4. Origins of climate finance |
|---|
| In the beginning of UNFCCC (1994), a financial mechanism was established to provide financial resources to developing country Parties. The operation of the financial mechanism can be entrusted to one or more existing international entities. |
| The Global Environment Facility (GEF) has served as an operating entity of the financial mechanism since the Convention's entry into force in 1994. |
| At COP 16 in 2010, Parties established the Green Climate Fund (GCF). In 2011, it was also designated as an operating entity of the financial mechanism. The financial mechanism is accountable to the COP, which decides on its policies, programme priorities, and eligibility criteria for funding. |
| In addition to providing guidance to the GEF and the GCF, Parties have established two special funds—the Special Climate Change Fund (SCCF) and the Least Developed Countries Fund (LDCF), both managed by the GEF—and the Adaptation Fund (AF) established under the Kyoto Protocol in 2001. |
| At the Paris Climate Change Conference in 2015, the Parties agreed that the operating entities of the financial mechanism – GCD and GEF – as well as the SCCF and the LDCF, shall serve the Paris Agreement. Regarding the Adaptation Fund serving the Paris Agreement, negotiations are underway in the Ad hoc Working Group on the Paris Agreement (APA). |
| Source: UNFCCC |

---

[12] https://unfccc.int/
[13] UNFCCC (2015). Paris Agreement, article 9.

The availability of resources for climate finance has been growing slowly for several years, below the financing pledges of developed economies. During the 15th Conference of Parties (COP15) of the UNFCCC in Copenhagen in 2009, developed countries committed to a collective goal of mobilizing USD 100 billion per year by 2020 for climate action with the goal of meaningful mitigation actions and transparency on implementation. The goal was formalized at COP16 in Cancun, and at COP21 in Paris, it was reiterated and extended to 2025. Currently, by the end of 2022, even though the recorded amount of finance flows reached a historic high, they are still below the needs and the amounts pledged. Moreover, climate finance resources are available, but they do not imply actual execution.

According to Climate Policy Initiative (CPI) (Buchner et al., 2023), "The global climate flow in 2021/2022 was USD 1.27 trillion. The cost of mitigating and adapting to climate change is estimated to range from USD 5.4 trillion to USD 11.7 trillion annually until 2030, and between USD 9.3 trillion and 12.2 trillion annually for the next two decades."

The CPI report found that in the average scenario, the annual climate finance needed through 2030 will increase steadily from $8.1 to $9 trillion. This amount will reach $10 trillion each year from 2031 to 2050. A fivefold increase in climate finance every year is necessary to prevent the worst impacts of climate change.

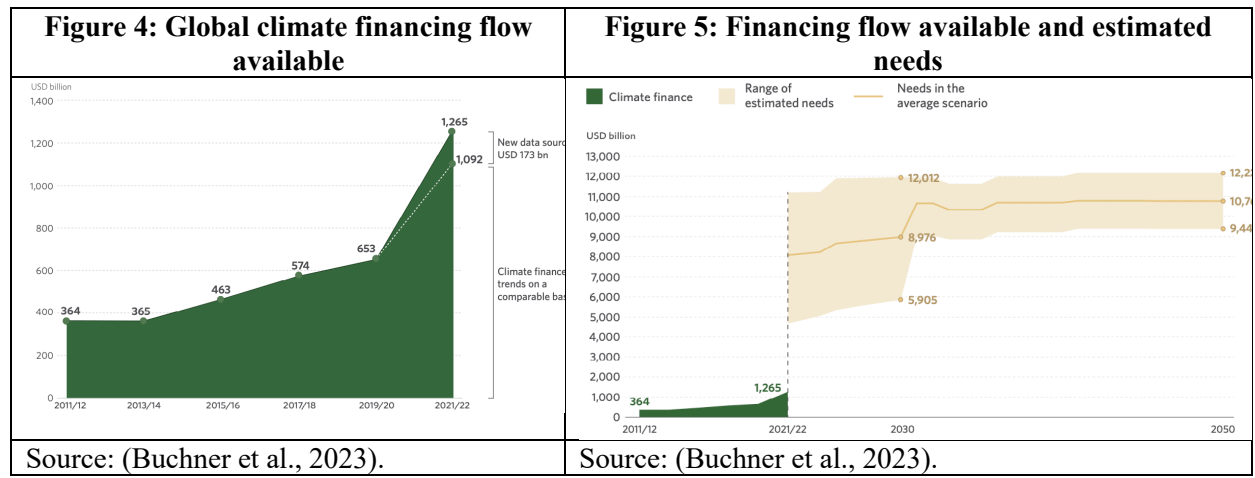

| Figure 4: Global climate financing flow available | Figure 5: Financing flow available and estimated needs |
|---|---|
| Source: (Buchner et al., 2023). | Source: (Buchner et al., 2023). |

Failure to achieve the goal of staying below the global warming trajectory of less than 1.5ºC implies higher costs for all economies in addition to the devasting costs due to disasters caused by climate-related natural phenomena. In addition, there would be a devastating impact on economic activity in terms of production capacity and the welfare of the world's population.

Some studies (Swiss Re, 2021) indicate that an increase in warming above 3ºC could lead to losses of 18% of global GDP by 2050 and 20% by 2100.

CPI estimates shows in Figure 6 how increasing climate investments to the levels needed by 2050 (USD 266 trillion cumulatively) will lead to a considerable reduction in social and economic losses by 2100: USD 1,266 trillion lower compared to a business-as-usual (BAU) scenario. In other words, sticking to BAU would create more than double the losses of a 1.5ºC scenario.

**Figure 6: Climate finance needs vs losses to be avoided**

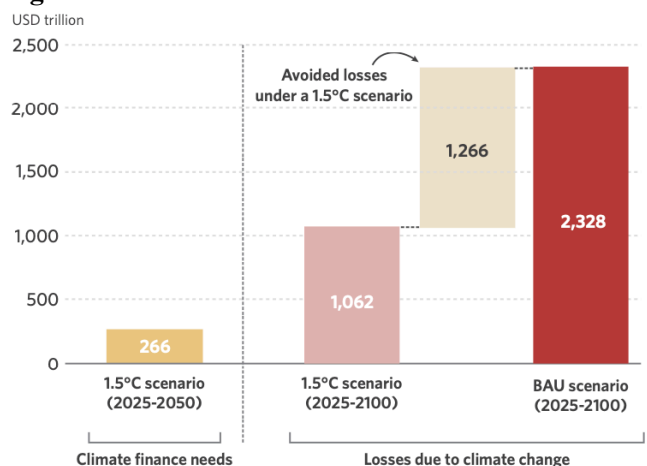

Source: CPI (2023)

## 4. Climate risk regulatory frameworks: individual countries efforts

Since the launch of the TCFD in December 2015 by the FSB, in G7 countries (European Union, Canada, France, Germany, Japan, England and the United States) concrete steps have been taken towards increased and better disclosure of financial and non-financial information of regulated companies by the financial authorities of each respective country.

On a global scale, there are other several economies that are considering this possibility but have not made concrete progress on drafting proposals for regulation in this matter.

In the chronology of each country (see Figure 7), the issuance of the TCFD was a watershed that catalyzed the drafting of various regulations among the G7 countries.

The regulatory frameworks analyzed propose their own mechanisms for requesting such information, as well as for disclosure to the financial markets. A common denominator among them is to modify, through the new regulatory frameworks, the reporting formats for inter-annual and annual reports, or to request to follow new report templates.

The information per se does not change the current global warming trend; however, by interpreting it, we can modify public policy decisions, company management and, in general, the direction of the economy as a whole.

Climate models and the various scenarios that have been developed are a fundamental part of the information needed to establish sound decision-making processes (Qin et al., 2013). For example, downscaled climate models can be used to refine projections from global climate models to finer scales (e.g., reflecting local climate hazards)(Flint & Flint, 2012). Downscaled projections are available to companies and can be used to identify climate hazards that may affect their assets, operations, work force, and supply chains.

The second goal of regulatory frameworks is the development of risk management models within companies that, in addition to considering traditional financial risks, incorporate methodologies to include climate risk.

When exploring the history of regulation of the financial sector, there are several examples of in the beginning of that regulation, how information was, though essential, was not enough. For example, in the banking sector, with the Basel regulation introduced by the end of the 80's decade, whose cornerstone is information reporting, the solvency model is the key to understanding the financial solvency of a banking institution. Another example is the Solvency II regulation for insurers[14], which provides a large amount of information, all to be interpreted according to the solvency model that each regulatory authority has approved.

Finally, a third goal that would be expected from regulatory frameworks is public policy on specifically what measures will be taken to maintain the stability of the financial system and the economy as a whole. Measures must evaluate specific emission levels of the regulated company and its supply chain. Assessing the capitalization levels to determinate the sufficiency to cover the risk exposed to climate change, and if necessary, increases in such capitalization levels. Finally, taxes on activities that contribute to reduce higher level of emissions.

In light of the three goals described above, the current effort by the G7 economies is focused on the first goal in its drafting phase, as well as on the collection of non-financial climate-related information. In many cases, the regulation is not yet in place. The second goal is still under discussion in most countries, and the third goal is a long-term project.

For many countries, TCFD regulation has been framed within existing environmental, social, and governance (ESG) regulations (Gargantini & Siri, 2022; Hösli & Weber, 2022). It is important to note that both have points in common. In particular, the environmental aspects overlap. For this reason, in several countries, the vehicle for incorporating TCFD recommendations has been to modify existing ESG regulations or link them to existing regulations (KPMG, 2023).

---

[14] The process began around 2005 as a fundamental review of existing insurance regulations, as Solvency I.

**Figure 7: Timeline of TCFD implementation in local regulatory frameworks**

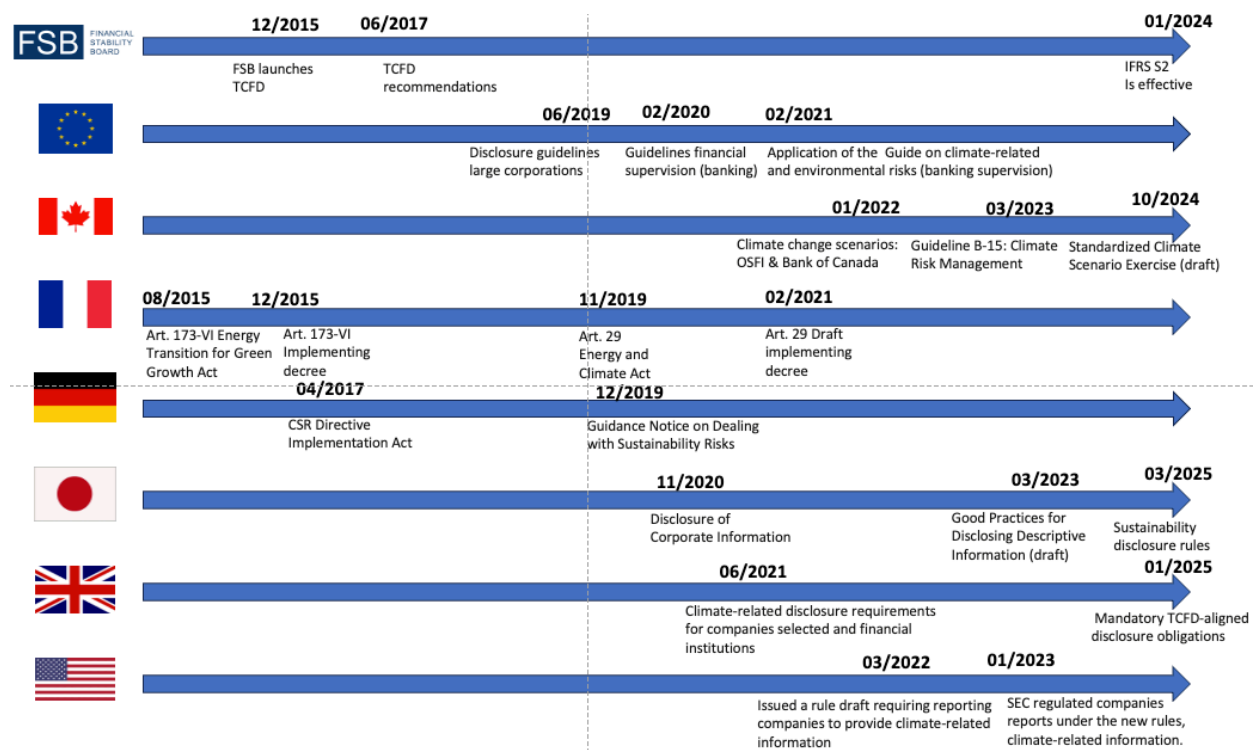

## 4.1. Current state of the regulatory framework on Canada peers' countries

*European Union*

The European Union issued its Action Plan on Sustainable Finance in 2019, which includes regulation on disclosures relating to sustainable investments and sustainability risks (3fP, 2019; CDSB, 2020).

Meanwhile, the European Banking Authority (EBA) published a draft implementing technical standards (ITS) on ESG risks under Pillar 3 of the Capital Requirements Regulation (CRR). As part of its climate-related and environmental risk management and disclosure guide, the European Central Bank (ECB) also issued supervisory expectations in 2020.

In March 2021, the Sustainability-related disclosures in the financial services sector (SFDR) took effect and was focused on financial institutions and financial advisers as a channel to investors.

As part of its review of the Non-financial Reporting Directive (NFRD) for large companies (including financial institutions), the Commission launched a public consultation in parallel. This led the Commission to propose a Corporate Sustainability Reporting Directive (CSRD) in April 2021, which would also apply to Small Medium Enterprises (SMEs)(Hilke et al., 2021).

In 2022, the publication of the European Commission Delegated Regulation (EU) 2022/1288 was released, which contained technical standards to be used by financial market participants when disclosing sustainability-related information under the Sustainable Finance Disclosures Regulation (SFDR). The requirements took effect on 1 January 2023 (Pietrancosta, 2022).

Finally, in September 2023, the European Commission launched 2 consultations through a public consultation on the implementation of the Sustainable Finance Disclosures Regulation (SFDR). Both consultations will end by the 15 of December, 2023.

*Canada*

The Canadian financial regulation has been making progress since 2015 in regard to climate-related discussions within the FSB as member of this organization. Several studies and assessments have been conducted by key relevant players showing a concern for the need of regulation on climate related risk (Bush & Lemmen, 2019; Galvez Rosa et al., 2022).

In the area of domestic regulation, efforts have been undertaken in two areas in the financial sector. The first has been the regulation of the securities market through the provincial regulator members of Canadian Securities Administrators (CSA)(SCO, 2019, 2021; TR, 2021). The second has been the regulation of federally regulated financial entities through the Office of the Superintendent of Financial Institutions (OSFI) (OSFI, 2023a).

In 2019, the CSA released Staff Notice 51-358 (SCO, 2019), Reporting of Climate Change-related Risks, assessing the main developments in relation to climate-related disclosures. In 2021, Canadian authorities published the proposed National Instrument 51-107 Disclosure of Climate-related Matters (SCO, 2021). As described in the proposal, the requirements would be phased-in over a one-year period for non-venture issuers and over a three-year period for venture issuers.

Regarding the federal regulated financial institutions regulated by OSFI, in March 2023, OSFI released a guideline for consultation on federally regulated financial institutions' management of climate-related risks. As part of the draft guideline, OSFI introduced mandatory climate-related financial disclosures that incorporate the TCFD recommendations. In October 2023, they released the Standardized Climate Scenario Exercise (draft for consultation)(OSFI, 2023b). They plan to release the final methodology for implementation by 2025.

*France*

Prior to the formal initiation of EU discussions, France played a pioneering role within the European Union. The inception of regulatory initiatives in France can be attributed to the 2015 Energy Transition Law (Hilke et al., 2021). This legislation mandated disclosure requirements for both institutional investors (Article 173 section VI) and companies (Article 173 section IV), necessitating the provision of information on climate change, environmental, social, and governance (ESG) factors, as well as their respective strategies and decisions (Baker & PRI, 2017).

The article 173 section IV and VI has been substituted in 2019 by the Energy and Climate Act article 29-II to become compatible with the Paris Agreement. This article includes activities of most financial players, including the banking sector.

*Germany*

In 2017, the CSR Directive Implementation Act (CSR-RUG) was instated (BJ, 2017), which relates to climate risk disclosure. In 2019, the Ministry of Finance issued the "Guidance Notice on Dealing with Sustainability Risk" with the aim to provide a guide on how to deal with sustainability risks to the entities supervised by the Federal Financial Supervisory Authority (Bundesanstalt für Finanzdienstleistungsaufsicht – BaFin) (BaFin, 2020).

The central focus of the Guidance Notice is risk management. It considers risk identification, management, and control processes together with traditional methods and procedures, with specific reference to sustainability risks. The main regulation linked to this Guidance Notice are KWG (Kreditwesengesetz – German Banking Act), VAG (Versicherungsaufsichtsgesetz – German Insurance Supervision Act) and KAGB (Kapitalanlagegesetzbuch - German Investment Code), regarding the inclusion of sustainability risks into risk management.

Another amended regulation was in 2020, the German Stock Corporation Act (AktG), section 161, adapting it in terms of the key thematic aspects required by TCFD recommendations, i.e.,in terms of corporate governance, strategy, and risk management (Leicht & Leicht, 2022).

The current regulatory framework in Germany is a mix of CSR-RUG, AktG and European Directives, as well as the German Accounting Standard (GAS), which complies with IFRS Foundation regulation.

*Japan*

In 2020, the Japanese financial authority, the Financial Service Agency, issued an Ordinance Concerning Disclosure of Corporate Information, following the TCFD recommendations (TFCD Japan, 2022). The proposed regulation in terms of reports issued by securities issuers includes a new section for "Sustainability-related philosophy and initiatives," "Governance" and "Risk management" are now required entries, and "Strategy" and "Indicators and targets" are important entries (FSA, 2022; Fumigada, 2022).

Based on the recommendations of the TCFD, the Bank of Japan released a report in 2022 describing Japanese climate change initiatives, as well as was developed an analytical exercise on climate-related risk for Japan (FSA & BJ, 2022).

The changes in the Japanese regulatory framework target the securities market, amending the Financial Instruments Exchange Act of Japan (FIEA). A new section, "Sustainability Perspective and Measures," was added to introduce ESG-specific disclosures for stock market companies in Japan (Schmittmann, 2022). These amendments are designed to address a company's sustainable development consistent with its management policies and strategies. The implementation of these changes should be completed by 2025.

The amendments will apply to the securities registration statement and annual securities report (Disclosure Documents), which are related to the fiscal year ending on and after March 31, 2023 (Kyle Lawless et al., 2023).

*United Kingdom*

The regulatory framework was developed in three sectors: corporations, pensions, and financial asset managers. Regarding commercial companies, the UK government introduced a new rule in Listing Rule 9.8 requiring that commercial companies with a UK premium listing (including sovereign-controlled commercial companies) disclose in their annual financial report whether they have made disclosures consistent with the recommendations of the TCFD. This regulation took effect in January 2021. In the same year, it was proposed to extend this regulation to securities issuers in stock markets, specifically to UK premium listing to issuers of standard listed equity shares (excluding standard listed investment entities and shell companies). The regulatory measures will be effective as of 2023 (FCA, 2021b, 2021a).

In relation to financial pension market participants, in 2021, the UK government issued a regulation requiring trustees with over £5 billion in relevant assets, as well as all master trust schemes and collective money purchase schemes authorized by the government, to engage in TCFD recommendations. Additionally, by October 2022, this regulation, previously only applicable to trustees with over £5 billion, became applicable to trustees with at least $1 billion.

Since 2022, financial assets managers with more than £50 billion in assets under management and asset owners (e.g., life insurers) with assets over £25 billion are required to make disclosures consistent with the TCFD recommendations on an annual basis at both an entity level and product level (FCA, 2020, 2021b, 2021a).

*United States*

In May 2021, the "Executive Order (EO) on Climate-Related Financial Risk " was released, consisting of six sections. The first three sections are highlighted below in relation to the possible changes in the regulatory framework that could be brought about (White House, 2021).

According the EO section 1, the federal government is recognizing its key role in managing the impacts of climate change through physical and transition risk inherent to assets, publicly traded securities, private investments, and companies. It also acknowledges its role in acting to mitigate those risks and its drivers. It states that "This policy will marshal the creativity, courage, and capital of the United States necessary to bolster the resilience of our rural and urban communities, States, Tribes, territories, and financial institutions in the face of the climate crisis…" (White House, 2021).

In section 2 of the EO, a key advisory body responsible for the development of the Climate-Related Financial Risk Strategy was established, which will be integrated by the Director of the National Economic Council and National Climate Advisor in coordination with the Secretary of the Treasury and the Director of the Office of Management and Budget. The strategy would include the following:

a) The measurement, assessment, mitigation, and disclosure of climate-related financial risk to Federal Government programs;
b) Financing needs associated with achieving net-zero greenhouse gas emissions for the U.S. economy by no later than 2050, limiting global average temperature rise to 1.5 degrees Celsius; and
c) Areas in which private and public investments can play complementary roles in meeting these financing needs.

In section 3, it is specified that the Financial Stability Oversight Council (FSOC) shall engage with FSOC members to develop an assessment of climate-related financial risk by financial regulators linked to section 1. Its primary objectives are to assess, in a detailed and comprehensive manner, the climate-related financial risk, including both physical and transition risks, to the financial stability of the Federal Government and the stability of the U.S. financial system. It must also facilitate the sharing of climate-related financial risk data and information among FSOC member agencies and other executive departments.

According to the FSOC Staff Progress Report on Climate-related Financial Risk in 2023 (FSOC, 2023), the FSOC is currently working as an interagency with regard to information sharing, coordination, and capacity building on climate-related financial risks. In this regard, the Joint Analysis Data Environment (JADE) was established to improve regulators' access to data, high-performance computing tools, and analytical and visualization software. By the end of 2023, JADE will support broad-based financial stability research by providing a platform to access and analyze a broad spectrum of financial and other relevant data and by facilitating data sharing and collaboration among Council member agencies.

By the end of 2022, the Department of the Treasury's Federal Insurance Office (FIO) issued a request for comment on a proposed collection of data from property and casualty insurers regarding current and historical underwriting data on homeowners' insurance at the ZIP Code level. The purpose was to assess the potential for major disruptions of private insurance coverage in regions of the country that are particularly vulnerable to the impacts of climate change.

The dedicated Council's Climate-related Financial Risk Committee is developing a climate scenario analysis (CSA) working group composed of Council member staff provides a forum for Council members—including the Federal Reserve, the Commodity Futures Trading Commission (CFTC), the Federal Deposit Insurance Corporation (FDIC), the Federal Housing Finance Agency (FHFA), and the Office of the Comptroller of the Currency (OCC) to explore the use of scenario analysis by regulators and regulated firms (FSOC, 2023).

The dedicated CFRC Risk Assessment Working Group (RAWG), which is composed of Council member staff, is developing a robust framework to identify and assess climate-related financial risk and is iteratively identifying a preliminary set of risk indicators for banking, insurance, and financial markets.

As part of the first changes in the regulatory arena, in March 2022, the Securities and Exchange Commission issued new requirements for reporting companies listed in stock markets to provide certain climate-related information in their registration statements and annual reports filed with the SEC (SEC, 2022).

The new requirements ask companies to include a new section in annual reports and registration statements titled "Climate-Related Disclosure," which would include climate-related governance, risk, and business impacts. They will also need to report on current greenhouse gas (GHG) emissions, together with an attestation report from an independent GHG emissions expert(CADWALADER, 2022; CG, 2022; Diamond et al., 2022). Finally, they must include a new note to a registrant's audited financial statements that provides climate-related metrics and impacts on a line-item basis. This set of regulations is fully aligned with TCFD recommendations.

**Other countries: current state of the regulatory framework**

Considering TCFD, CDP and CDSP (CDP & CDSB, 2019; TCFD, 2022; TFCD, 2023) there are other countries involved in the analysis and implementation of their own regulatory frameworks :
- In 2021, the Australian Prudential Regulation Authority (APRA) published guidance for banks, insurers, and superannuation trustees for managing financial risks associated with climate change.
- In 2021 in Brazil, the stock market regulator amended its rules to require securities issuers to indicate whether they disclose environmental, social, and corporate governance, considering TCFD recommendations.
- In 2021, the Egyptian Financial Regulatory Authority announced the issuance of resolutions requiring companies listed on the Egyptian Stock Exchange and companies operating in non-bank financial activities to submit disclosure reports related to sustainability and the financial impacts of climate change in line with the TCFD recommendations.

- In 2021, the Hong Kong government issued an official document through the Mandatory Provident Fund Schemes Authority with high-level principles for mandatory provident fund trustees on integrating ESG factors into their investment and risk management processes. The same year, the Hong Kong Monetary Authority issued a supervisory policy manual for banks, restricted license banks, and deposit-taking companies (authorized institutions) on key elements of managing climate-related risk.
- In 2022, the Reserve Bank of India (RBI) released a Discussion Paper on Climate Risk and Sustainable Finance to seek feedback on several topics, including climate-related financial disclosure.
- In 2022, the Malaysian government released a guide to support implementation of climate-related disclosures aligned with TCFD recommendations. The guide is targeted to the Bank Negara Malaysia and the Securities Commission Malaysia and includes commercial banks, investment banks, insurance and reinsurance companies, and fund management companies.
- In 2021, the New Zealand Government passed legislation inspired by TCFD with regard to making climate-related disclosures mandatory for large publicly listed companies, insurers, banks, non-bank deposit takers, and investment managers.
- In 2022, the Swiss Federal Council developed an "implementing ordinance" on climate reporting for large Swiss companies; this ordinance was motivated by TCFD recommendation. Thailand is still in the process of assessing potential changes in its regulatory framework. For example, the Bank of Thailand issued a consultation paper on the financial landscape that describes policies to support the financial sector, including management of environmental risks.
- Finally, in 2023, Mexico's local representation of IFRS proposed a new set of norms for companies inspired by IFRS S1 and S2 to be followed by market players. The authorities are still in the process of tailoring financial regulations.

## 5. Analysis of the regulatory framework of climate risk for the financial sector in Canada

The aim of this section is twofold. Firstly to consider the regulatory framework issued by the federal authority (OSFI) for regulated financial entities. The analysis is based on evaluating whether there is adherence to the international standard issued by IFRS through S2 standards. Then, to explore how that adherence relates to the international standard.

While some of the regulation is still under development, part of the new regulation will be designed considering an analytical exercise. to be commented by financial market players This report analyzes existing information about that exercise as of November 15, 2023. Additionally, this analysis considers the OSFI: Guideline on "Climate Risk Management (CRM)" (OSFI, 2023a) issued in March 2023, the analytical exercise was released as "Standardized Climate Scenario Exercise (SCSE)" (OSFI, 2023b) issued in October 2023.

The methodology of this report identifies opportunities within the guideline issued and the analytical exercise currently in the process of generating feedback. The key to assessing the impact of climate change on the financial entities regulated is to evaluate how this regulation will close the gap between the current knowledge of the impacts of climate change and the knowledge that is required to mitigate and/or adapt to climate impacts.

Considering that the IFRS S2 (ISSB, 2023a) took over the lead on the TCFD recommendations, the following subsections will assess each theme considering the content of IFRS S2 in connection with Canadian regulation.

| **IFRS S2 standard** |
|---|
| **Objective**<br>The objective of IFRS S2 Climate-related Disclosures is to require an entity to disclose information about its climate-related risks and opportunities that is useful to primary users of general-purpose financial reports in making decisions relating to providing resources to the entity.<br><br>**Scope**<br><br>IFRS S2 applies to:<br><br>1. Climate-related risks to which the entity is exposed, which are:<br>   1.1. climate-related physical risks; and<br>   1.2. climate-related transition risks; and<br>2. Climate-related opportunities available to the entity.<br><br>Climate-related risks and opportunities that could not reasonably be expected to affect an entity's prospects are outside the scope of this Standard.<br><br>**Objectives by theme**<br><br>- **Governance**<br><br>The objective of climate-related financial disclosures on governance is to enable users of general-purpose financial reports to understand the governance processes, controls, and procedures an entity uses to monitor, manage, and oversee climate-related risks and opportunities.<br><br>- **Strategy**<br><br>The objective of climate-related financial disclosures on strategy is to enable users of general-purpose financial reports to understand an entity's strategy for managing climate-related risks and opportunities.<br><br>- **Risk management**<br><br>The objective of climate-related financial disclosures on risk management is to enable users of general-purpose financial reports to understand an entity's processes to identify, assess, prioritise, and monitor climate-related risks and opportunities, including whether and how those processes are integrated into and inform the entity's overall risk management process.<br><br>- **Metrics and targets**<br><br>The objective of climate-related financial disclosures on metrics and targets is to enable users of general-purpose financial reports to understand an entity's performance in relation to its climate-related risks and opportunities, including progress towards any climate-related targets it has set, and any targets it is required to meet by law or regulation. |
| *Source: Taken from IFRS S2, IFRS Foundation (2023)* (ISSB, 2023a). |

## 5.1. Analysis of the regulatory framework

Canada's regulatory framework for Federal Regulated Financial Entities (FRFI's) in the context of climate change and related climate risks has been evolving, from its first approximations in discussions among regulators within FSB in 2015 on the first TFCD findings in 2016, to now in 2024 having 7 initiatives that seek to implement the original recommendations of TFCD and IFRS S2 at national level.

These initiatives, which aim to 'support FRFIs in their efforts to build awareness and capability in managing climate- related financial risks. Improving FRFI readiness to manage climate-related financial risks enhances the safety and soundness of these institutions and strengthens public confidence in Canada's financial system.'

The initiatives consider a broad spectrum of climate-related risk issues, including a general guide for risk management, the development of a database of relevant information describing climate risks, a scenario-based approach to analysis, and an analysis of potential impacts on institutions' capitalization and liquidity levels. A first exercise on the type of information to be disclosed to the authorities and the market, consistent with international initiatives. A strategy to involve relevant stakeholders link to regulated entities. Finally, a strategy to increase the capacities of the financial authorities and regulations, which is why they have created a forum to constantly deepen the various technical aspects linked to climate risk analysis.

In particular, the content of the latest version of the CRM guide, released in March 2023, was analyzed. For its analysis, it was contrasted with the international framework issued by IFRS S2.

In Annex 8, each section is analyzed in relation to the international framework. It was found to be highly consistent with the objective, scope and core content of the international framework (IFRS S2).

Regarding governance and risk management, the CRM guide is based on 6 principles, which additionally contemplate issues related to these topics, as well as specific aspects of scenario analysis, capital adequacy and liquidity levels.

Regarding the disclosure of information, it is contained in 6 main principles, which seek that the information be relevant, comprehensive, specific, reliable, verifiable, appropriate in terms of the characteristics of the issuing entity in terms of size, nature and complexity, and finally consistent over time.

The CRM guide aims to develop the first attempts to implement it in 2024, 2025 and 2026 and will be voluntary.

*Summary of findings on SCSE*

There is significant consistency between the characteristics of the SCSE and similar exercises carried out by Canadian G7 peers, as well as the recommendations of international regulatory organizations regarding the estimation and analysis of climate risks. Thus, we believe the analysis is focused on examining key aspects of current estimation methodologies and application of the analysis that is focused on areas of the Canadian economy that are crucial to its growth and net zero emission goal.

As was stated by OSFI in the SCSE climate scenario analysis is in its nascency and in general, the financial methodologies, particularly risk management techniques for financial entities, are yet to be updated to include climate considerations formally. However, the state of knowledge in finance and climate science in Canada is at the cutting edge of science among peers.

In our view, SCSE is a well-supported exercise for the short term; however, it is limited by construction, because, for example, it does not consider the same level of detail in physical risk as it does for transition risk. In this way, the current exercise is achievable in the short term, as was proposed by the authority in 2025. However, many key challenges remain, especially regarding physical risk. This should be implemented in the medium to long-term. For example, for the United States analytical exercise, it has defined a broader strategy by using a deeper scope of analysis, similar to Canada's, but including aspects currently not included in SCSE, such as consideration of physical risk (FSOC, 2023; White House, 2021).

Although the Climate Risk Hub has been established as a forum that aims to be inclusive, attracting a variety of sectors, including academia and financial sector participants, there is currently no known work strategy with a specific action plan that engages, calls, and follows up on the meetings with the various interlocutors. In certain OSFI announcements, it has been noted that agreements have been reached with participants, including climate risk modelers, to enhance OSFI's capabilities. However, these appear to be isolated instances; there is no known general action plan for universities and research centers to collaborate with defined objectives and deliverables.

For example an action plan could include themes as follow:1) the selection and calibration of loss functions suitable for Canadian exposure, 2) the financial modeling dynamic in financial statements through tailor-made methods to the constraints of data availability, and 3) suggestions on the selection on the types of events to be classified low recurrency losses (high return periods) high recurrency losses (low return periods). This selection could shape the final outcome of the loss estimation. The impact of climate change may be more evident in high-return periods.

Therefore, as a result of its new climate-related risk regulation (OSFI, 2023a), Canada is closer to its G7 peers than it has ever been. However, the progress is still insufficient to propose concrete and specific measures to manage climate risk for financial institutions. As of now, the financial system's supervisory authority has proposed an exercise of estimating climate risk and disclosing non-financial information. This exercise of estimation is biased towards transition risk, except for flooding. Yet, it is unclear when and how the analysis could be extended to include physical risk in the same detail as transitional risk or compounding risk. In light of the current state of climate risk quantification, changes in capitalization levels in institutions or creating additional reserves, as many others measures to be assessed are a far-off task for the time being, given the current state of climate risk quantification.

*Specific comments on SCSE*

This report can help inform future studies when it comes to evaluating Canada's financial stability and the goal to achieve net-zero emissions. We suggest further reflection on a few aspects:

- A bias in two directions is observed. The analysis sought by the SCSE seems biased towards banks, with a particular emphasis on transition risks rather than physical risks.
    - For example, the estimation of climate indicators considering climate risk such as ECL, PD, and LGD, whereas the SCSE did not consider physical risk.
    - In the case of insurance companies, it is not clear how the current exercise includes analytical outcomes estimating physical risk.

- When analyzing within the same financial institution by sector, it is not clear what priority should be given to those sectors that are less represented in the income of financial institutions (e.g., SMEs).

- There is no clarity about how this assessment should be developed by size of entities, for small entities could imply a heavy regulatory charge. On the other hand, large entities may require sophisticated modeling assessments using state-of-the-art techniques.

    - In the international experience, some regulators propose a "mock of model" relevant for both the smallest financial entities and for the largest ones.

- It would be beneficial to provide more details on the workbook and its instructions before formally releasing it.

### 5.2. Main findings in Canadian Federal proposed regulation on Financial Entities

After assessing the proposed regulatory framework (CRM guide and SCSE), we would like to highlight the following items and comments:

- The guideline complies with international practices in general terms; however, it creates similar concerns, specifically with regard to how it should be implemented by size, sector, and data granularity of each financial institution.
- The Canadian financial regulator confirms its leadership among G7 peers in terms of the regulatory proposals.
- Although the regulatory framework proposed (guideline on Climate Risk Management) generally complies with topics among G7 peer economies, Canada and its peers are facing a lot of challenges implementing a relevant, suitable, and useful regulation to reach net zero emissions goals. In this regard, SCSE is a crucial exercise to assess the impact of climate change in the financial sector.
- The risk transition approach by the SCSE is generally consistent with previous assessments, such as the ones developed by the The Bank of Canada and OSFI in 2022 (BoC & OSFI, 2022). The literature suggests that this approach is consistent with international experience of similar assessments. However, this approach leaves open several windows of opportunity, such as those highlighted by (Monasterolo et al., 2023):
    - Integrating physical risk within transition risk, since neither occurs independently.
    - Embedding the role of investors' expectations and climate in the scenarios' trajectories, among others.
- The physical risk approach could be expanded in terms of depth. For instance, it could assess the implication of physical risk in terms of profitability, creditworthiness, and solvency of financial entities.
- The SCSE project must be rendered compatible with other assessments (flooding risk assessment) as a single analytical effort, considering uniformed metrics in the assessment, and explained to the financial community.
- The OSFI has indicated that it does not attempt to make concrete estimates of risk by "size the risk" (OSFI, 2023b). We agree with this strategy as the short-term strategy stated in the CRM and SCSE. However, they may be improved in the medium term through more in-depth analyses and by organizing working groups within Canadian universities to fill the knowledge gap. More research must be done on damage functions selection, financial modeling, and optimizing exceedance probability curves, among other topics.

# 6. Canada's challenges to achieve smooth energy transition supported by the financial sector

Canada is confronted with a dual position in the energy transition. Firstly, it serves as a regulator of the sectors that are integral to the energy value chain. Particularly, the financial sector. Simultaneously, it derives advantages from the energy sector that it endeavors to diminish, specifically fossil fuel energy. From this source, it obtains royalties that significantly contribute to its public expenditure. Additionally, Canada is a producer in a market with clear incentives to continue growing, particularly in the fossil fuel energy sector, which is highly competitive and dynamic. This defines the three primary challenges of the energy transition, which are provided in the following section.

*Challenge 1: strengthening the financial regulatory framework*

The complexity of climate change is scaled to the financial sector. This sector is still in a nascent stage to finance the transition to a sustainable economy, achieving an economy with zero net emissions is a major challenge. At the same time, the financial system may be impacted by the sector's current exposure to brown financing and questionable sustainability of those financial assets in investor's portfolios in the future.

Canada is one of the top ten countries emitting $CO_2$, with a goal to reduce emissions to 500 million tonnes. Currently, Canada emits approximately 730 million tonnes of $CO_2$. A study by RBC (RBC, 2022) estimates that a 75% reduction in emissions by 2050 would imply mobilizing close to $2 trillion over three decades. This estimation implies the government, businesses and communities will have to spend annually roughly $60 billion USD, this investment is significant, as it involves a substantial leap from the current $15 billion. This investment aims to support the development of new technologies and may be further enhanced by implementing natural solutions, such as extensive tree planting and the preservation of wetlands and grasslands.

In terms of policy making, the financial sector could catalyze investments and reduce emissions. However, there is a debate among scientists, practitioners and regulators, as well as leaders in the financial industry, about how far financial regulation should go to address climate change concerns. In Levonian (2024), the author highlights the view of some regulators in this regard, for instance the Federal Reserve Board Chair Jerome Powell (2023), "The Fed does have narrow, but important, responsibilities regarding climate-related financial risks …tightly linked to our responsibilities for bank supervision…. But without explicit … legislation, it would be inappropriate for us to use our monetary policy or supervisory tools to promote a greener economy or to achieve other climate-based goals. We are not, and will not be, a
'climate policymaker.'"

However, other financial authorities (US Treasury, European Banking Authority, European Securities and Markets Authority, German Federal Ministry for Economic Affairs and Climate Action (BMWK), Banque de France, among others) are currently discussing how regulation could discourage investment in assets linked to high emissions and promote investment in those that are sustainable.

Two instruments have been discussed on regulating financial climate risk, capital requirements for regulated entities, which could help in risk coverage, as a temporary buffers, for instance supervisors may require banks to hold additional capital as a buffer to compensate for potential understated expected losses due to climate change. This buffer is seen as a temporary measure until methodologies for capturing expected losses are more developed. Capital requirements provide a structure for addressing capital adequacy and includes mechanisms for adjusting risk weights for different types of exposures, as well as guide the public policy to promote a financial sector that finances a sustainable economy or to restring the capital to high emission investments. Another instrument could be through concentration Limits, based on relevant metrics of climate risk exposure, such as carbon emissions, can be a more direct approach to limiting risk than relying solely on capital requirements.

The implementation of such regulations, which are currently under discussion, allows regulators to moderate the risk tolerance of brown and green assets. As a result, market discipline could respond by directing the behavior of financial market participants, including depositors, shareholders, and securities purchasers, who monitor and influence the behavior of financial institutions. Under this model, participants are highly motivated to effectively manage risks and generate revenue. Financial institutions and borrowers are expected to respond to signals from regulators and financial markets in order to preserve their solvency and stability. This procedure may involve the establishment of prices, the modification of business practices, and the compliance with regulatory requirements.

In conclusion, the regulatory framework is an essential component of financial climate risk management. This is precisely why an ambitious scope of the regulator's studies on financial climate risk estimation, which takes into account the most recent trends in scientific research, can generate substantial benefits for the economy, the financial sector, and the public sector. Since it enables the analysis of the feasibility of strategies to achieve the defined goals in relation to emissions reduction based on scientific principles.

*Challenge 2: Decarbonize the Canadian public finance*

The oil and gas industry has long been a cornerstone of the Canadian economy, significantly influencing the fiscal system. Over the past decades, the contribution of oil revenues to Canada's Gross Domestic Product (GDP) has been substantial, impacting both federal and provincial fiscal policies.

As of 2021, the oil and gas extraction industry accounted for an average of 5% of Canada's GDP, 21% of Alberta's GDP, and 25% of Newfoundland and Labrador's GDP since 2000. This sector's economic significance is underscored by its substantial contribution to employment, investment, and exports.

Over the past decade, the value of Canadian crude oil exports has seen a dramatic increase, rising more than fifteen-fold and accounting for 14.1% of Canada's total exports in 2019.

Therefore, oil revenues have been particularly crucial for oil-producing provinces like Alberta and Newfoundland and Labrador. In Alberta, the oil sector's contribution to GDP has been a major driver of provincial revenues.

At the federal level, oil revenues contribute to the national fiscal framework through corporate taxes, royalties, and other levies. The federal government has used these revenues to fund public services and infrastructure projects.

Therefore, meeting the target and reducing emissions may have an impact on tax revenues, which poses the challenge of finding new sources of revenue to replace the financing needs of the government at the federal, provincial and local levels.

*Challenge 3:  Invest decisions toward a net zero emissions goal*

The Canadian energy regulator has defined certain trajectories for achieving emission levels and refocus to renewal as energy source by 2050, instead of oil and gas-based energy. The development of these energy sources imposes investment challenges.

Canada is currently financing energy transition to the tune of approximately USD 15 billion. This amount is expected to rise to USD 80 billion by 2030 and some analyst estimate that should reach around USD 2 trillion by 2050.

There are two perspectives that may be helpful in framing the decision about investing in an energy project: the first is the developer's viewpoint, which stresses the demand for funding, and the second is the investor's viewpoint, which highlights the supply side of funding, in which the investor enables the project to be financed.

According to the first viewpoint, the cost of capital is one critical measure. In the 2023 report from the Energy Transition Risk and Cost of Capital Programme (ETRC) provides a comprehensive analysis of the cost of capital trends in the global energy sector, with a particular focus on electric utilities and energy production. The report expands its scope to include the cost of equity in addition to the cost of debt, utilizing data from various sources including corporate bonds, the Platts World Electric Power Plants Database (WEPP), and the Institutional Brokers' Estimate System (IBES).

- o Electric Utilities: the global trends of renewable electric utilities consistently show a lower cost of capital compared to their fossil fuel counterparts. For instance, the cost of debt for renewable electric utilities is around 6%, compared to 6.7% for fossil fuel utilities. Similarly, the cost of equity is lower for renewables (15.2% vs. 16.4% for fossil fuels).

- o Energy Production: the global trends of coal mining has the highest cost of capital within the energy sector, with significant increases since the Paris Agreement in 2015. Renewable fuels & technology have a lower cost of capital than oil & gas production but higher than oil & gas services.

In the energy production, as shown in the graph (left) below for debt, the cost of capital in oil fuel projects are higher than renewable projects around the world.

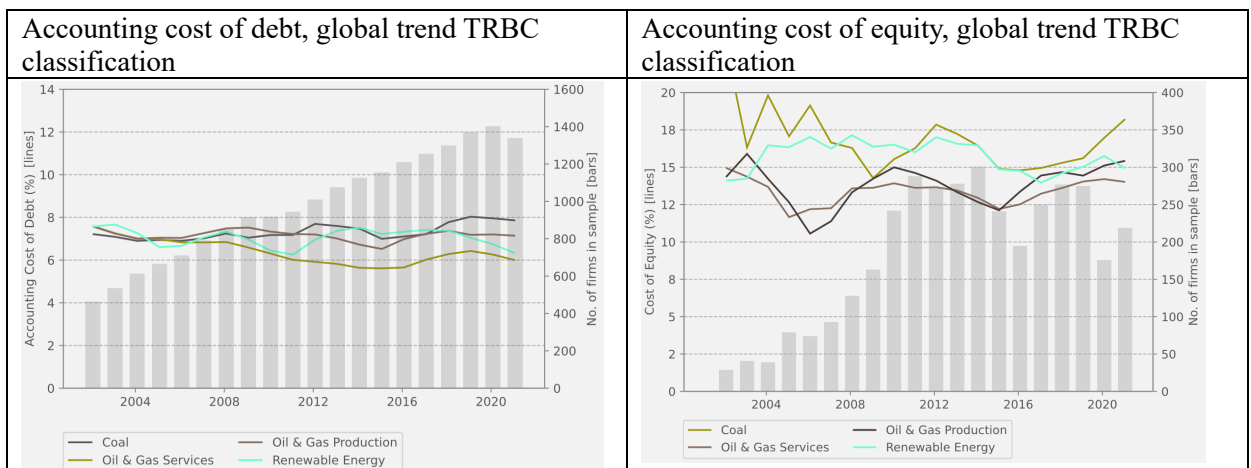

| Accounting cost of debt, global trend TRBC classification | Accounting cost of equity, global trend TRBC classification |

Source: Zhou et al (2023)
/* Refinitiv Business Classification (TRBC) sector classification. TRBC sector classification was originally developed by the Reuters Group and has been owned by Refinitiv since 2018. It is the basis for Refinitiv Indices.

On the other hand, according to IEA during the last two decades there was an astonishing increasing in the compound annual growth rate of 23%, the yearly investment flows accessible for renewable energy projects (see below). This shows that the pattern in the expansion of investment supply of resources underlines a significant shift globally.

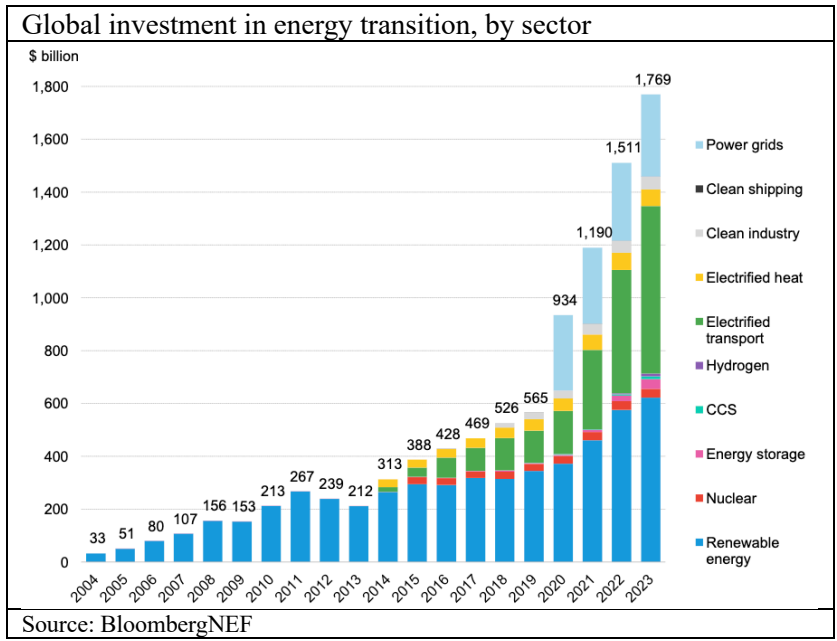

Regarding the supply side of investment, a key indicator is the return of the investment, EROI, or Energy Return on Investment, is a key indicator used to assess the efficiency and sustainability of energy sources. It measures the amount of energy that is returned from an energy-generating process relative to the amount of energy invested to obtain that energy. In other words, EROI indicates us how much energy we get back for every unit of energy we put in to extract or produce energy. An EROI greater than 1 indicates that the energy source is providing more energy than what is required to produce it, leading to a net energy gain. Conversely, an EROI less than 1 means that the energy source is not providing enough energy to cover the costs of its production, resulting in a net energy loss.

Assessing EROI by different energy sources:

- o Conventional Fossil Fuels: These include oil, coal, and natural gas. Historically, conventional fossil fuels have had high EROI values, but as the most accessible and highest-quality reserves are depleted, their EROI is declining. For example, the EROI of conventional oil has decreased over time as more energy is required to extract oil from deeper wells or from more challenging geological formations.

- o Unconventional Fossil Fuels: These include shale oil, tar sands (oil sands), and deepwater oil. Unconventional fossil fuels generally have lower EROI values compared to conventional ones because they require more energy-intensive processes to extract and refine. For instance, the extraction of oil from tar sands requires significant energy inputs for mining and upgrading the bitumen to synthetic crude oil.

- Renewable Energy Sources: These include solar, wind, hydroelectric, and geothermal energy. The EROI of renewable energy sources can vary widely depending on the technology and the specific location. For example, solar photovoltaic (PV) systems have seen a significant increase in their EROI due to technological advancements and economies of scale. Wind energy, especially in favorable locations, can have very high EROI values, often exceeding those of conventional fossil fuels.

- Nuclear Energy: Nuclear power plants have relatively high EROI values, but the calculation can be complex due to the unique characteristics of nuclear fuel cycles and the long-term management of nuclear waste. The EROI for nuclear energy can also vary based on the type of reactor and the efficiency of the fuel cycle.

According to Delannoy et al (2021), indicates that *"the energy necessary for the production of oil liquids is currently about 15.5% of the gross energy produced and is expected to grow exponentially, reaching 50% by 2050"* This energy is used directly and indirectly in the form of fuel for machinery, electricity for operations, and other material inputs necessary for the oil production process.

The "gross energy produced" refers to the total amount of energy that is extracted from oil liquids before any energy is consumed in the production process. The "energy necessary for production" is the portion of this gross energy that must be reinvested back into the production process to maintain or increase oil output.

The study projects that this energy reinvestment will grow exponentially over time, reaching a point where it will consume 50% of the gross energy produced by 2050. This implies that by the middle of the century, half of the energy produced from oil liquids will be needed just to keep the production process going. This growth in energy reinvestment is a concerning trend because it leaves less net energy available for other uses, such as powering the economy, supporting societal functions, or facilitating a transition to low-carbon energy sources.

In summary, for the energy industry, particularly the fossil fuel-based industry, there are dark clouds on the horizon, raising questions about a smooth and easy energy transition for Canada. Precisely because the industry, on which a considerable part of its public spending depends, is profitable and is expected to remain profitable until at least 2050.

In general, the renewable energy industries continue to encounter obstacles in enhancing their profitability in order to become more competitive, despite the consistent reporting of higher EROIs.

Some believe this may happen sooner than anticipated or already is happening. For example, Brockway et al (2019) estimated final-stage energy-return-on-investment (EROI) ratios for fossil fuels are significantly lower than those at the primary energy stage. According to the document, the global fossil fuel EROI at the final energy stage (EROIFIN) was around 6:1 in 2011, which is a substantial decrease from the primary energy stage EROI (EROIPRIM) of approximately 30:1. This indicates that the energy return on investment for fossil fuels at the final stage of energy consumption is much closer to the EROI ratios of renewable energy sources than previously expected.

For comparison, the EROI ratios for renewable energy sources, particularly electricity generated from photovoltaics and wind turbines, are typically estimated to be in the range of 5:1 to 20:1 at the final energy stage. This suggests that the transition to renewable energy sources may not result in a significant reduction in economy-wide EROI, as the final-stage EROI ratios of fossil fuels are already relatively low and declining.

In spite of the findings of Brockway et al (2019), investment in fossil fuel projects continues to draw new investment, despite the global discourse on climate change and emissions reductions.

## 7. Conclusions

- The Canadian financial regulatory framework for climate risk management is consistent with international practices but faces challenges in implementation, particularly in terms of data granularity and sector-specific requirements.

- There is a need for further reflection on the balance between transition risks and physical risks in the regulatory framework, with a current bias towards banks and transition risks.

- The document suggests that the regulatory framework should be expanded to include comprehensive disclosure requirements that are relevant, reliable, and consistent over time.

- The implementation of the CRM guide is planned to be voluntary in the process to be implemented, concluding in 2026, indicating a gradual approach to adoption.

- The document highlights the importance of climate scenarios for the financial sector, emphasizing the need for further study to ensure proper application in valuation and analysis.

- Shareholders remain concerned about potential changes in capital requirements based on climate risk, as the methodology for estimating climate risk is still under development.

- The document underscores the significance of systematic and orderly disclosure of non-financial information, which is crucial for effective climate risk management.

- The regulatory frameworks of various countries propose mechanisms for requesting and disclosing climate-related information, often modifying reporting formats to include such disclosures.

- Climate models and scenarios are essential for sound decision-making processes, helping to identify potential climate hazards affecting assets, operations, and supply chains.

- The Canadian financial system has significant exposure to the oil and gas industry, which poses challenges for transitioning to a low-emission economy while maintaining financial stability and economic growth.

- In the document identifies three challenges for Canada:
    - Challenge 1: Strengthening the Regulatory Framework. Develop a robust regulatory framework to guide the financial sector in managing climate risks and supporting the transition to a sustainable economy.

    - Challenge 2: Decarbonizing Canadian Public Finance. The oil and gas industry has been a cornerstone of the Canadian economy, significantly influencing fiscal policies. Meeting emission reduction targets may impact tax revenues, requiring new sources of revenue to replace the financing needs of the government.

- Challenge 3: Investment Decisions Toward a Net Zero Emissions Goal. The Canadian energy regulator has defined trajectories for achieving emission levels and refocusing on renewables by 2050, which demands new investments. However, in general, the renewable energy investment continues to encounter obstacles in enhancing their return on the investment in order to become more competitive, despite the consistent reporting of higher EROIs. Meanwhile, return on investment of oil and gas industry remains high, at least until 2050, as is expected.

## 8. Annex

### 8.1. Governance

The following analysis compares the IFRS S2 norm by theme with the Canadian regulation, in terms of the CRM and SCSE. Paragraphs (P) 1 to 4 correspond to the objectives and scope.

| Reference in IFRS S2 | Considered by Canadian Regulation | Reference in Canadian Regulation | | | Comment |
|---|---|---|---|---|---|
| | | CRM | SCSE | other | |
| P5 | ✓ | Chapter I Principle 1 & 2 | | OSFI's Corporate Governance Guideline, Section I (OSFI, 2018) | |
| P6 (a) from (i) to (v) | ✓ | Chapter I Principle 1 & 2 | | OSFI's Corporate Governance Guideline, Section II, IV & V (OSFI, 2018) | |
| P6 (b) from (i) to (ii) | ✓ | Chapter I Principle 1 & 2 | | OSFI's Corporate Governance Guideline, Section II (OSFI, 2018) | |
| P7 | ✓ | Chapter I Principle 1 | | | The aim of the regulation is climate related risk disclosure without duplication with Sustainability-related Financial Information (IFRS S1) |

### 8.2. Strategy

| Reference in IFRS S2 | Considered by Canadian Regulation | Reference in Canadian Regulation | | | Comment |
|---|---|---|---|---|---|
| | | CRM | SCSE | Other | |
| P8 | ✓ | A3-1, Chapter I section I | | | |
| P9 from (a) to (e) | ✓ | A3-1, Chapter I section I & Chapter I section III | | OSFI Guideline E-18: Stress Testing (OSFI, 2009) | |
| P10 from (a) to (d) | ✓ | Chapter 1, section I Principle 2, and section II, subsection A | | | |
| P11 | ✓ | Chapter 1, section I Principle 2, and section II, subsection A.7 | | | |
| P12 | ✓ | Chapter 1, section I Principle 2, and section II, subsection A | | | |
| P13 from (a) to (b) | ✓ | Chapter 1, section I Principle 2, and section II, subsection B | | | There is no explicit mention of the entity's business model linked to value chain climate-related risk exposure |
| P14 (a) from (i) to (v) | ✓ | Chapter 2, section III | | | |
| P14 (b) | ✓ | Chapter 2, section III | | | |
| P14 (c) | ✓ | Chapter 2, section III | | | |
| P15 from (a) to (b) | ✓ | Chapter 2, section III, Principle 1, Chapter 1, section IV | | | |
| P16 from (a) to (b) | ✓ | Chapter 2, section III, Principle 1 | | | |
| P16 (c) from (i) to (ii) | ✓ | Chapter 2, section III, Principle 1 and Principle 2 | | | |
| P16 (d) | ✓ | Chapter 2, section III, Principle 1 and Principle 2 | | | |
| P17 | ✓ | Chapter 2, section III, | | | |
| P18 from (a) to (b) | ✓ | Chapter 2, section III, Principle 3 | | | |
| P19 from (a) to (b) | ✓ | Chapter 2, section III, Principle 1 | | | |
| P20 | ✓ | Overview A1 | | | |
| P21 from (a) to (c) | ✓ | Overview A1 | | | |
| P22 (a) from (i) to (iii) | ✓ | Chapter 1, section III | | SCSE section 2.2 | After the exercise, OSFI expects to finalize the methodology and issue the SCSE in 2024. |
| P22 (b) from (i) to (iii) | ✓ | Chapter 1, section III | | SCSE section 2.2 | |
| P23 | ✓ | Annex 2-2 | | SCSE section 3.3 | |

## 8.3. Risk Management

| Reference in IFRS S2 | Considered by Canadian Regulation | Reference in Canadian Regulation | | | Comment |
|---|---|---|---|---|---|
| | | CRM | SCSE | Other | |
| P24 | ✓ | Chapter 1, section II | | | |
| P25 (a) from (i) to (vi) | ✓ | Chapter 1, section II, subsection A | | | |
| P25 (b) | ✓ | Chapter 1, section II, subsection B | | | |
| P25 (c) | ✓ | Chapter 1, section II, subsection A & B | | | |
| P26 | ✓ | Overview A1 | | | |

### 8.4. Metrics and targets

| Reference in IFRS S2 | Considered by Canadian Regulation | Reference in Canadian Regulation | | | Comment |
|---|---|---|---|---|---|
| | | CRM | SCSE | Other | |
| P27 | ✓ | Annex 2-1 and 2-2 | | | |
| P28 from (a) to (c) | ✓ | Annex 2-1 and 2-2 | | | |
| P29 (a) from (i) to (vi) | ✓ | Annex 2-1 and 2-2 | | | |
| P29 (b) | ✓ | Overview A3 Outcomes, Chapter 2 Principle 2 | | | No explicit mention |
| P29 (c) | ✓ | Overview A3 Outcomes, Chapter 2 Principle 2 | | | No explicit mention |
| P29 (d) | ✓ | Overview A3 Outcomes, Chapter 2 Principle 2 | | | No explicit mention |
| P29 (e) | ✓ | Overview A3 Outcomes, Chapter 2 Principle 2 | | | No explicit mention |
| P29 (f) from (i) to (ii) | ✓ | Annex 2-1 | | | No explicit mention |
| P29 (g) from (i) to (ii) | ✓ | | | OSFI's Corporate Governance Guideline, Section II, "FSF Principles for Sound Compensation Practices" | No explicit mention |
| P30 | ✓ | Overview A1 | | | |
| P31 | | Chapter 2, Section III Principle 2, Annex 2-2 | | | |
| P32 | ✓ | Overview A1 | | | |
| P33 from (a) to (h) | ✓ | Chapter 1, Section II-B | | Guideline E-19: Own Risk and Solvency Assessment Guideline E-19: Own Risk and Solvency Assessment | |
| P34 from (a) to (d) | ✓ | Chapter 1, Section II-B | | | |
| P35 | ✓ | Chapter 1, Section II-B | | | |
| P36 from (a) to (d) | ✓ | Annex 2-2 | | | |
| P36 (e) from (i) to (iv) | ✓ | Annex 2-2 | | | |
| P37 | ✓ | Overview A1 | | | |

### 8.5. Assessment on the SCSE

#### 8.5.1. Check list of the SCSE

| Reference in IFRS S2 Appendix B Application guidance | Considered by Canadian Regulation | Reference in Canadian Regulation | | | Comment |
|---|---|---|---|---|---|
| | | CRM | SCSE | Other | |
| B1 | ✓ | Chapter 1 section III | Section 2.1, 2.2 | | |
| B2 | ✓ | Chapter 1 section III | Section 2.2 | | |
| B3 | ✓ | Chapter 1 section III | Section 2.1 | | |
| B4 | ✓ | Chapter 1 section III | Section 2.2 | | |
| B5 | ✓ | Chapter 1 section III | Section 2.1 | | |
| B6 | ✓ | Chapter 1 section III | Section 2.1 | | |
| B7 | ✓ | Chapter 1 section III | Section 2 | | |
| B8 | ✓ | Chapter 1 section III | Section 2.2.2 | | |
| B9 | ✓ | Chapter 1 section III | Section 2.2.1 | | |
| B10 | ✓ | Chapter 1 section III | Section 2.2.2 | | |
| B11 | ✓ | Chapter 1 section III | Section 3.2.2, 4 and 5 | | |
| B12 | ✓ | Chapter 1 section III | Section 3.2.2 and 5.2 | | |
| B13 | ✓ | Chapter 1 section III | Section 2.1 and 2.2 | | |
| B14 | ✓ | Chapter 1 section III | Section 2.1 and 2.2 | | The SCSE is above the international standard |
| B15 | ✓ | Chapter 1 section III | Section 2.2.1., 2.2.2 | | |
| B16 | ✓ | Chapter 1 section III | Section 2 | | |
| B17 | ✓ | Chapter 1 section III | Section 2.1 | | |
| B18 | ✓ | Chapter 1 section III | Section 2 Overview | | |

### 8.5.2. Description of the SCSE: a summary

In October 2023, the OSFI released its comments on the SCSE. The document contains six sections, including one for appendices. The aim of the document is to describe an exercise as a first trial to assess climate risk. The document assumes that climate risk is the result of interaction of physical and transition risk, and both could have a *significant impact in soundness of financial institutions and the Canadian financial system*.

The aim of this exercise is to get feedback before the methodology is officially released in 2024. It will include a set of instructions, workbook, and questionnaire.

According to the SCSE, the focus of the exercise is on risk discrimination and exposure assessment given the complexity of climate change, the financial sector, and the uncertainty associated with climate in the future. *The OSFI believes this discrimination is achievable between counterparties, industries and financial entities regulated.* There are three objectives:

1) Raising awareness and encouraging a strategic orientation with FRFIs to better understand the potential exposures to climate change.
2) Encouraging the building of FRFIs' capacity to assess the impact of climate related catastrophic events and policies and to conduct climate scenario analysis exercises.
3) Establishing a standardized quantitative assessment of climate related risks, both transitional and physical in nature.

This assessment is the first step for the OSFI in climate risk assessment. The outcome will shape future exercises. For now, SCSE exercise will not pretend to be a comprehensive sizing of climate risk.

Regarding the operation approach, the exercise will be performed in dual directions. From top-down, the "*OSFI will define and develop the 2024 SCSE methodology, scenarios, adjustment parameters, and calculations.*" From bottom-up, the "*FRFIs will identify exposures, classify them into relevant sectoral and geographical segments, and perform calculations.*" The implementation will be executed in three steps: *1) OSFI develops the SCSE methodology, 2) OSFI prescribes scenarios, risk parameters, formulas, etc. to FRFIs , OSFI prescribes scenarios, risk parameters, formulas, etc. to FRFIs, and 3) FRFIs assess impacts to their exposures using prescribed information from OSFI.*

This will be an independent and uncorrelated analysis. It is projected to be developed in four different modules.

| Scope | Climate Risk | Exposures | Financial Risk |
|---|---|---|---|
| Impact of Climate Transition on Market Risks for Commercial Exposure | Transition Risk | Commercial (Global) | Market Risk |
| Impact of Climate Transition on Credit Risk for commercial Exposures | Transition Risk | Commercial (Global) | Credit Risk |
| Climate Transition Real Estate Exposure Assessments | Transition Risk | Real Estate-related (Canadian) | Exposure Assessment |
| Physical Risk Exposure Assessments | Physical Risk | Real Estate-related (Canadian) | Exposure Assessment |

**Transition risk**

The analysis for transition risk will consider transition risk drivers, transition channels, and risk parameters used in credit and market risk assessments, according to the following table:

| **Transition risk drivers** | **Transition channels** | **Risk parameters used in credit and market risk assessments** |
|---|---|---|
| • Government climate policies | **Microeconomics** | • Net income/earnings |
| • Technological change | • Households | • Firm asset value/enterprise value |
| • Changes in consumer preferences | • Corporates | • Risk free rates |
|  | • Issuer specific financial assets | • Corporate credit spreads |
|  | **Macroeconomics** |  |
|  | • Overall economy |  |
|  | • Macroeconomic variables |  |

The analysis would resemble the one developed by the Bank of Canada/OSFI 2021.

**Scenarios for transition risk analysis**

The scenarios would be taken from those developed by the Bank of Canada scenario data prepared by the Bank of Canada and customized to the Canadian economy. This analysis will use scenario data developed by the Network for the Greening of the Financial System (NGFS), the international standard setter for climate scenarios data, specifically NGFS Phase III scenarios.

**Industry sector classification in transition risk analysis**

The transition risk analysis will include 25 identified industry sectors, which includes electricity production, energy intensive industries, fossil fuels, transportation, agricultural forestry, and other sectors. It will also assess the real estate transition risk exposure.

Regarding the industry transition risk analysis, for the lending firms regulated, a study will be conducted on the credit risk impact, considering the base scenario of the following indicators: Expected Credit Loss (ECL)[15], probability to default (PD)[16] and Loss Given Default (LGD)[17]. For each indicator the analysis proposes to estimate a climate adjustment for ECL[18], PD[19], and LGD[20], according to an equation described in the SCSE.

In terms of the impact of climate in market risk, according to the SCSE, *climate transition risk could significantly impact the value of financial assets such as stocks and bonds. The scope of exposures for the market risk module includes equities and corporate bonds in both the trading and banking books. The asset classes that are in scope for the credit risk module are: corporate debt and shares.* This analysis will assess equity shocks and corporate credits spreads. Both assessments should consider different dimensions such as climate scenario narrative, year, industry sectors and geography.

**Real estate transition risk exposure assessment**

According to the SCSE, the aim of this module is not to attempt to measure financial impact. OSFI has identified possible transmission channels related to the transition away from a carbon-intensive economy that may impact the risks associated with real estate lending and investment: *1) Exposures to properties that are powered or heated by carbon-intensive sources such as fossil fuels or natural gas may be impacted by the transition to net-zero. 2) Borrowers employed in industries exposed to higher transition risks may face additional financial hardship given shifts in the labour market.* The scenarios will be taken from the same used for transition risk analysis, and on the other hand, the exposure assessed would be highly relayed on mortgages sector as well as data from for small and medium enterprise, and corporates.

---

[15] ECL is the probability-weighted estimate of credit losses (i.e., the present value of all cash shortfalls) over the expected life of a Financial Instrument. A cash shortfall is the difference between the cash flows that are due to an entity in accordance with a contract (the scheduled or contractual cashflows) and the cash flows that the entity expects to receive (the actual expected cashflows). Given that expected credit losses consider both the amount and timing of payments, a credit loss arises even if the entity expects to be paid in full but later than when contractually due. (Taken from https://www.openriskmanual.org/wiki/Expected_Credit_Loss).

[16] PD is the probability of an obligor fails in the payment of a debt or financial commitment documented in a financial instrument.

[17] LGD is the amount estimated of loss by a financial entity when a borrower default on a loan.

[18] $\text{climateECL} = \sum_{k=1}^{m} w_k \sum_{i=0}^{n} \text{climatePD}_i^k \times \text{climateLGD}_i^k \times \text{EAD}_i^k \times \frac{1}{(1+r)^i}$

[19] $\text{climatePD}_i^k = \frac{1}{1 + \exp(-(\text{logit}(\text{PD}_i^k) + \text{climateAdd-on}_i))}$

[20] $\text{climateLGD}_i^k = \frac{\Phi[\Phi^{-1}(\text{climatePD}_i^k) - \Phi^{-1}(\text{PD}_i^k) + \Phi^{-1}(\text{PD}_i^k \times \text{LGD}_i^k)]}{\text{PD}_i^k}$

**Physical risk**

OSFI recognizes that physical risk is a source of losses for regulated financial institutions, and that chronic or acute threats can cause such losses. The proposed analytical exercise does not seek to estimate such financial losses the same way it does for transition risk.

This analysis will take into account the Representative Concentration Pathways (RCP) scenarios, specifically 2.6, 4.5 and 8.5.

The hazards to be considered are high temperatures, flooding from rivers and rain, as well as wildfire.

The exposure will be described by: retail mortgages, home equity line of credit, small and medium enterprises and corporates, as well as other collateralized commercial lending exposures that are in the scope of the physical risk exposure assessment. The database of exposure should be georeferenced (i.e., latitude and longitude).

The physical risk analysis should be developed for real estate and commercial lending exposures.

The physical risk exposure assessment is segmented by the following dimensions:
- Province
- Asset class (as listed in Section 5.4) Asset class (as listed in Section 5.4)
- LTV buckets (for real estate exposures only)
- Credit quality buckets, if available,
    - For individual borrowers, credit score buckets
    - For corporate borrowers, credit rating buckets
- Physical hazard buckets

The exposure assessment will include the following aggregated amounts The exposure assessment will include the following aggregated amounts
- Balances outstanding Balances outstanding
- Authorized amounts
- Weighted average PD
- Weighted average LGD
- Insurance liabilities (if applicable)